\documentclass[aps,prb,twocolumn,floatfix]{revtex4-2}
\usepackage{graphicx}
\usepackage[colorlinks=true,linkcolor=blue,citecolor=blue,urlcolor=blue]{hyperref}
\usepackage{siunitx}
\DeclareSIUnit\rydberg{Ry}
\DeclareSIUnit\atomicunit{a.u.}
\DeclareSIUnit\bohr{\text{\ensuremath{a_0}}}
\usepackage[all]{hypcap}
\usepackage{float}
\usepackage{amsmath}
\usepackage{amssymb}
\usepackage{xcolor}
\usepackage{xr}

\newcommand{\minus}{\text{--}}

\newcommand{\VEC}[1]{\mathbf{#1}}

\begin{document}

\title{CrTe$_2$ as a two-dimensional material for topological magnetism in complex heterobilayers}

\author{Nihad Abuawwad$^{1,2,3,*}$, Manuel dos Santos Dias$^{4}$, Hazem Abusara$^{3}$, Samir Lounis$^{1,2.*}$}
\address{$^1$ Peter Gr\"unberg Institut and Institute for Advanced Simulation, Forschungszentrum J\"ulich \& JARA, 52425 J\"ulich, Germany}
\address{$^2$ Faculty of Physics, University of Duisburg-Essen and CENIDE, 47053 Duisburg, Germany}
\address{$^3$ Department of Physics, Birzeit University, PO Box 14, Birzeit, Palestine}
\address{$^4$ Scientific Computing Department, STFC Daresbury Laboratory, Warrington WA4 4AD, United Kingdom}
\date{\today}%

\begin{abstract}
The discovery of two-dimensional (2D) van der Waals magnetic materials and their heterostructures provided an exciting platform for emerging phenomena with intriguing implications in information technology.
Here, based on a multiscale modelling approach that combines first-principles calculations and a Heisenberg model, we demonstrate that interfacing a CrTe$_2$ layer with various Te-based layers enables the control of the magnetic exchange and  Dzyaloshinskii-Moriya interactions as well as the magnetic anisotropy energy of the whole heterobilayer, and thereby the emergence of topological magnetic phases such as skyrmions and antiferromagnetic N\'eel merons. The latter are novel particles in the world of topological magnetism since they arise in a frustrated N\'eel magnetic environment and manifest as multiples of intertwined hexamer-textures. Our findings pave a promising road for proximity-induced engineering of both ferromagnetic and long-sought antiferromagnetic chiral objects in the very same 2D material, which is appealing for new information technology devices employing quantum materials.
\end{abstract}

\maketitle

\section{Introduction}

Since the experimental demonstration of two-dimensional (2D) van der Waals (vdWs) magnets with intrinsic magnetism in 2017~\cite{cri3, cr2ge2te2}, research interest in 2D magnetic nanomaterials has grown rapidly due to their potential applications in spintronic devices and their significance in fundamental physical studies~\cite{Gibertini2019, Gong2019, McGuire2020, Jiang2021b, Sierra2021}. 
The possibility of heterostructuring offers the unprecedented possibility to engineer quantum materials with exquisite properties facilitated by the quasi-perfect interfaces expected by the vdW gaps. This enables the realization of totally new heterostructures hosting novel functionalities that are not otherwise seen in the individual building blocks. 
In most cases, these 2D materials have a simple collinear magnetic order, such as ferromagnetic, or antiferromagnetic. However, they can also exhibit complex noncollinear magnetism and host ferromagnetic skyrmions, to give just two examples.
Skyrmions are topologically-protected chiral spin textures of great interest for potential applications as information carriers in information technology devices~\cite{bog, nag, fert}. 
These chiral magnetic states typically arise due to the interplay between Heisenberg exchange and relativistic Dzyaloshinskii-Moriya interaction (DMI)~\cite{Moriya, dzy} in materials that lack inversion symmetry and have non-zero spin-orbit coupling.
In a 2D material, skyrmions arise when the magnetization points out of the plane of the magnetic layer.
When the magnetization is in-plane it instead gives rise to merons, which represent another form of chiral spin-textures.
The topological charge for skyrmions is integer, while it it half-integer for merons, hence both kinds of spin textures are qualitatively distinct~\cite{Goebel2021}.

In the context of 2D heterostructures, ferromagnetic (FM) Néel-type skyrmions were experimentally detected in  Fe$_3$GeTe$_2$~\cite{sky-0, sky-1, sky-2, sky-3, sky-4}, while FM merons were only evidenced in thin films and disks~\cite{m1, m2, m4, m5}.
Theoretically, a free standing monolayer of  CrCl$_{3}$ has been predicted to host such half-integer spin-textures, which are stabilized by the magnetic dipolar coupling that favors an overall in-plane orientation of the magnetization~\cite{m3}.
Another 2D material which is attracting interest due to its high magnetic ordering temperature is CrTe$_2$.
Recent studies have shown that a thin CrTe$_2$ layer grown either on SiO$_2$/Si or heterobilayer graphene substrates is ferromagnetic with a high Curie temperature of \SI{200}{\kelvin}~\cite{fm-crte2,Zhang2021}.
In contrast, a different study on a single monolayer of CrTe$_2$ deposited on graphene indicated  a zig-zag antiferromagnetic ground state while  a magnetic field drives a transition to a non-collinear spin texture~\cite{afm-crte2}.
We found that it exhibits magnetic frustration and strong magnetoelastic coupling, which depending on the resulting magnetic state leads to breaking of inversion symmetry and the emergence of DMI~\cite{Abuawwad}.
However, no magnetic skyrmions have been found in this material.

\begin{figure}[tb]
\includegraphics[width=\columnwidth]{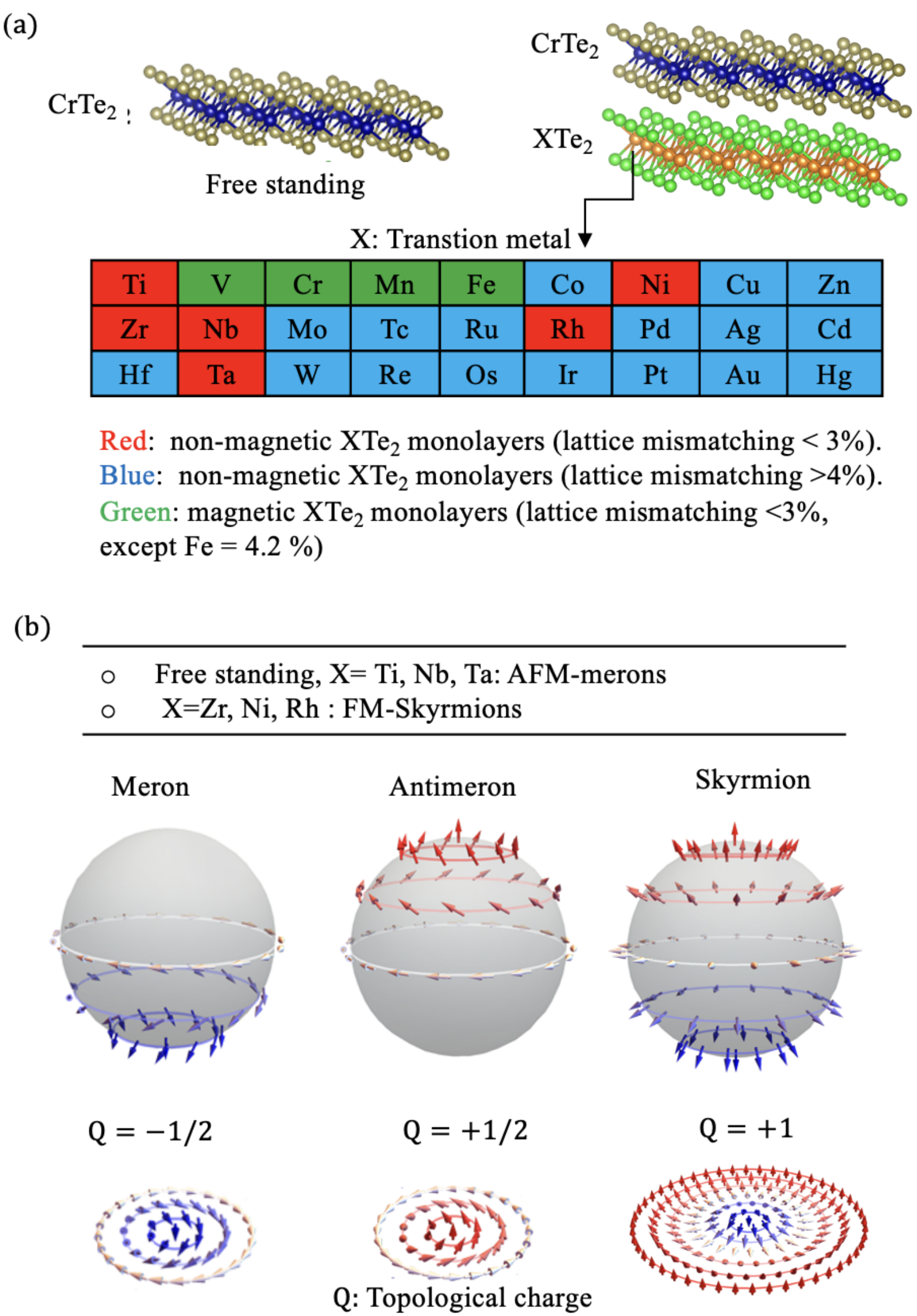}
\caption{\label{figure_1} {Overview of the heterostructures and magnetic states.
    (a) CrTe$_{2}$/XTe$_{2}$ heterobilayer, where X is the transition metal from the right table. (b) The topological magnetic states found in the CrTe$_{2}$/XTe$_{2}$ heterobilayer, with X being the transition metal in red color. The spin-textures of a meron, antimeron and skyrmion are illustrated together with the wrapping of the surface of a unit sphere by the underlying magnetic moments, which gives rise to different topological charges.}
   }
\end{figure}

Here, we explore the possibility of engineering 2D topological magnetism in a CrTe$_2$ monolayer by constructing heterostructures with Te-based layers involving other non-magnetic transition metal atoms (see Fig.~\ref{figure_1}).
Utilizing a multi-pronged approach based on first-principles calculations combined with an extended Heisenberg model, we unveil here new topological antiferromagnetic (AFM) objects already arising in the free-standing 1T phase of CrTe$_{2}$.
These objects consist of multi-meronic particles emerging in a frustrated in-plane N\'eel magnetic environment.
Such AFM topological states have long been sought in the context of skyrmionics as ideal information carriers, since they are expected to be unaffected by the skyrmion Hall effect~\cite{hall-1, hall-2, hall-3, hall-4, sky-halleffect-1, sky-hall-2, nag} responsible for the undesired deflection of conventional skyrmions from a straight trajectory upon application of a current.
Their AFM nature should also lead to a weak sensitivity to external magnetic fields and potentially terahertz dynamics~\cite{tera, tera-1}, further motivating efforts towards their experimental realization.
Once interfaced with various Te-based layers containing either heavy of light transition metal atoms, we demonstrate the ability to engineer the stability and nature of the underlying magnetic state.
Surprisingly, with the right vdW heterostructure, the AFM merons can be converted to FM skyrmions, which opens unique opportunities for designing devices made of 2D materials to realize fundamental concepts for information technology based on topological magnetic bits.

\section{Methods}

\subsection{First-principles calculations}

Structural relaxations of CrTe$_{2}$/XTe$_{2}$ heterobilayer were assessed using density functional theory (DFT) as implemented in the Quantum Espresso (QE) computational package \cite{qe} with projector augmented plane wave (PAW) pseudopotentials \cite{ps}.
In our calculations, the generalized gradient approximation (GGA) of Perdew-Burke-Ernzerhof (PBE)\cite{PhysRevLett.77.3865} was used as the exchange and correlation functional.
The plane-wave energy cut-off is \SI{80}{\rydberg}, and the convergence criterion for the total energy is set to \SI{0.01}{\micro\rydberg}.
We included a vacuum region of \SI{30}{\angstrom} in the direction normal to the plane of the heterobilayer to minimize the interaction between the periodic images. 
The residual forces on the relaxed atomic positions were smaller than \SI{1}{\milli\rydberg\per\bohr}, and the pressure on the unit cell smaller than \SI{0.5}{\kilo\bar}. 
The self-consistent calculations were performed with a k-mesh of $24 \times 24 \times 1$ points and the Brillouin zone summations used a Gaussian smearing of \SI{0.01}{\rydberg}.

Once the geometries of the various collinear magnetic states were established, we explored in detail magnetic properties and interactions with the all-electron full-potential relativistic Korringa-Kohn-Rostoker Green function (KKR-GF) method as implemented in the JuKKR computational package \cite{Papanikolaou2002,Bauer2014}\footnote{The juKKR code can be found at \url{https://jukkr.fz-juelich.de/}}.  
The angular momentum expansion of the Green function was truncated at $\ell_\mathrm{max} = 3$ with a k-mesh of $48 \times 48 \times 1$ points.
The energy integrations were performed including a Fermi-Dirac smearing of \SI{502.78}{\kelvin}, and the local spin-density approximation was employed~\cite{Vosko1980}.
The Heisenberg exchange interactions and Dzyaloshinskii–Moriya (DM) vectors were extracted using the infinitesimal rotation method \cite{inf-rot} with a finer k-mesh of $200\times 200 \times 1$.

\subsection{Magnetic interactions and atomistic spin dynamics}
The magnetic interactions obtained from the first-principles calculations are used to parameterize the following classical extended Heisenberg Hamiltonian with unit spins, $|\mathbf{S}| = 1$, which includes the Heisenberg exchange coupling ($J$), the DMI ($D$), the magnetic anisotropy energy ($K$), and the Zeeman term ($B$):

\begin{equation}\label{eq:spin_model}
E = -\sum_i \VEC{B}\cdot \VEC{S}_i - \sum_i K_i (S_i^z)^2
- \sum_{i,j} J_{ij} \VEC{S}_i \cdot \VEC{S}_j - \sum_{i,j} \VEC{D}_{ij}\cdot(\VEC{S}_i \times \VEC{S}_j).
\end{equation}

Here $i$ and $j$ label different magnetic sites within a unit cell.
The magnetic properties pertaining to CrTe$_{2}$ were evaluated by analysing the Fourier-transformed magnetic interactions, which in reciprocal space gives access to the magnetic ground state and the related dispersion of potential spin spirals: $J_{ij}(\VEC{q}) = \sum_j J_{0j}e^{-i\VEC{q}\cdot\VEC{R_j}}$, 
where $\VEC{R}_{0j}$ is a vector connecting unit cells atom 0 and $j$. 

Furthermore, atomistic spin dynamic simulations using the Landau-Lifshitz-equation (LLG) as implemented in the Spirit code \cite{Mueller2019a}\footnote{The Spirit code can be found at \url{https://juspin.de}} are performed in order to explore potential complex magnetic states while the geodesic nudged elastic band (GNEB) is utilized for the investigation of stability against thermal fluctuations \cite{genb, genb-1, genb-2}.

\section{Results and discussion}

\subsection{New topological AFM magnetic state in monolayer CrTe$_{2}$}

\begin{figure}[H]
\includegraphics[width=\columnwidth]{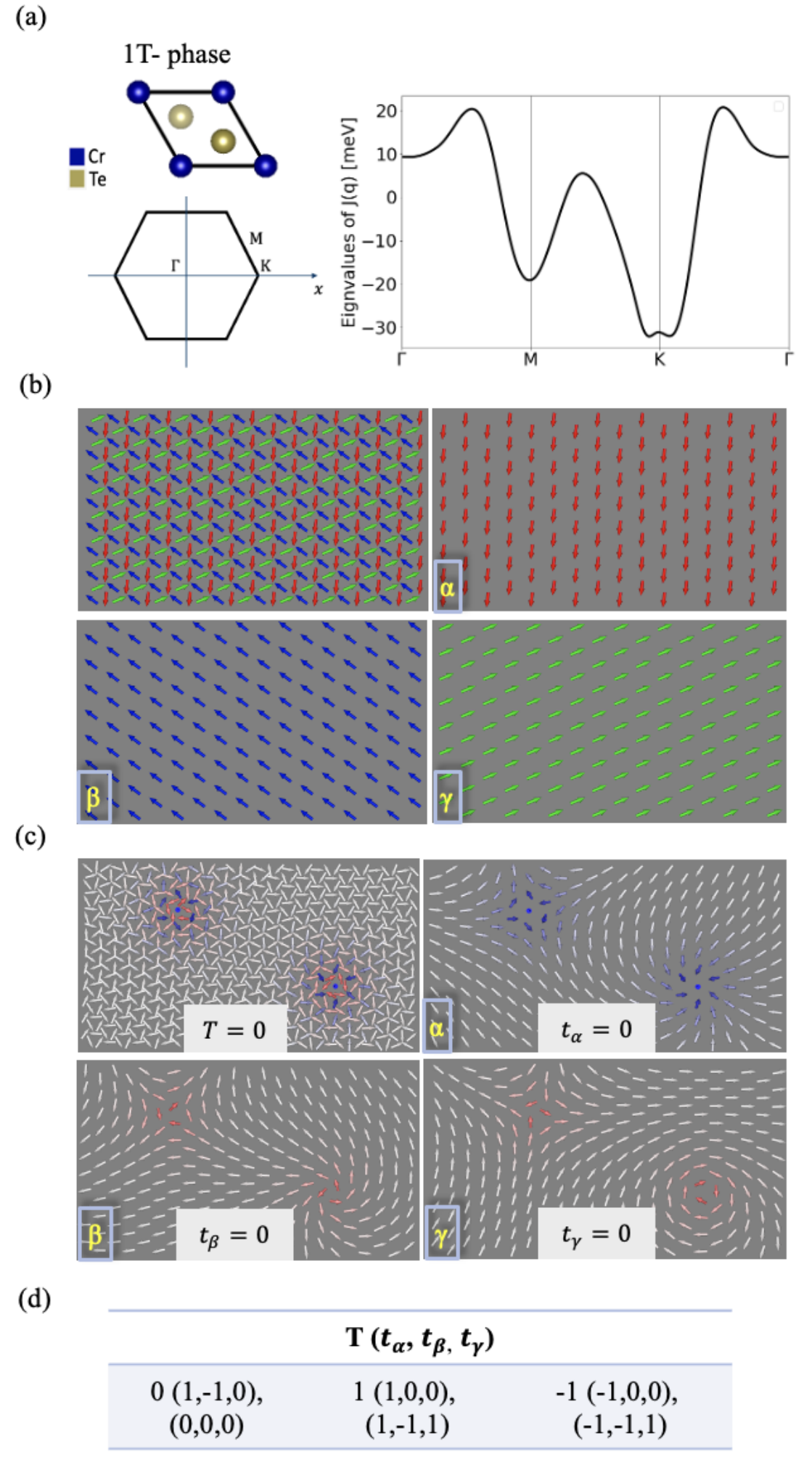}
   \caption{(a) The unit cell utilized for the simulation of the 1T phase of the free-standing monolayer of CrTe$_2$ (top), the hexagonal first Brillouin zone (bottom),  and eigenvalues of the Fourier-transformed exchange interactions as a function of $q$ in in
  free standing of CrTe$_2$ (right). 
   (b) Decomposition of the N\'eel state into three ferromagnetic sublattices $\alpha$, $\beta$, and $\gamma$ carrying moments rotated by 120$^\circ$.
   (b) Frustrated AFM multi-meronic spin-textures.
   The total topological charge $Q$ is decomposed into the three sublattices, each of which hosts a pair of merons with sublattice-dependent topological charges $t_i$ ($i = \alpha, \beta, \gamma$).
   Here the magnetic background is the spin spiraling ground state, which is very close to the N\'eel state.
   (d) Total topological charge $T$ and how it arises from various possible combinations of the topological charges from each sublattice.}
    \label{figure_2}
\end{figure}

The 1T phase of free-standing CrTe$_2$ monolayer is characterized by a magnetic moment of $2.67\, \mu_\mathrm{B}$.
The magnetic interactions favor antiferromagnetism, which on a triangular lattice usually leads to the N\'eel phase in which neighboring magnetic moments have an angle of 120$^\circ$. 
It can be partitioned into three FM sublattices named $\alpha$, $\beta$, and $\gamma$ as shown in Fig.~\ref{figure_2}(b). 
However, the Heisenberg exchange interactions are long-ranged and introduce competing tendencies, resulting in a frustrated spin spiraling state with an energy minimum close to the N\'eel state (Fig.~\ref{figure_2}(a)).
This spin spiral is further modified by the magnetic anisotropy (1.4 meV), which favors an in-plane orientation for the magnetic moment, and finally leads to the discovery of a new topological AFM state made of a hexamer of meronic texture.

In order to understand the origin of this magnetic phase, we show in Fig.~\ref{figure_2}(c) that the new topological AFM state arises from various combinations of meronic textures coexisting as pairs (meron-meron, meron-antimeron  and antimeron-antimeron) in each of the three AFM sublattices.
We note that within each sublattice, these pairs are ferromagnetic. 
The topological charge ($t$) for a meron is determined by the product $pw/2$, where $w$ is the
winding number $w$, which describes the in-plane rotation of the magnetic moments with $w=+1$ ($\minus 1$) for vortex (antivortex).
The polarity $p$ describes the out-of-plane core magnetization ($p=+1$ for up, and $p= \minus 1$ for down).
This leads to a value of $t= \minus 1/2$ ($t = + 1/2$) for a  meron (antimeron).
Various combinations of meronic states can emerge, leading to a rich set of the possible values of the total topological charge $T$ illustrated in Fig.~\ref{figure_2}(d). 
The first scenario illustrated in Fig.~\ref{figure_2}(c) corresponds to a meronic hexamer with zero total topological charge ($T = 0$), which can arise either when each sublattice accommodates a meron-antimeron pair ($t = 0$), or when hosting pairs of  meron-meron pair ($t = 1$), antimeron-antimeron ($t = \minus 1$) and  meron-antimeron ($t = 0$). 
The second and third scenarios have an opposite total topological charge of $+1$ and $-1$. The state with  $T = 1$ ($-1$) occurs either when two sublattices have meron-antimeron pairs carrying a charge $t = 0$ and third sublattice contains an antimeron-antimeron pair of a charge $t = 1$ (meron-meron pair with $t = -1$), or when two sublattices have antimeron-antimeron pairs with $t = 1$ (meron-meron pairs with $t = -1$) while the remaining sublattice host a meron-meron pair of charge $t = -1$ (antimeron-antimeron with $t = 1$).

\begin{figure}[htb]
\includegraphics[width=\columnwidth]{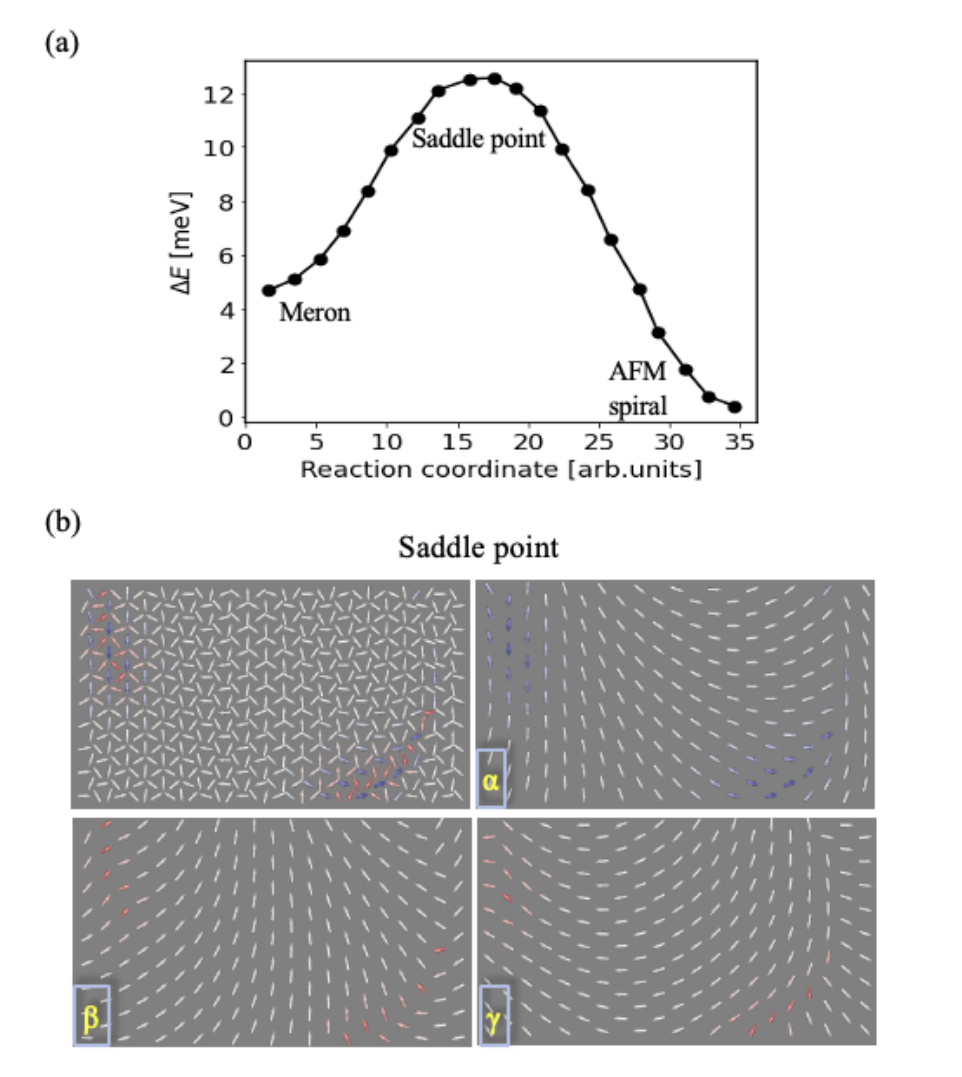}
  \caption{
 (a) Energy path for the collapse of the single pair of the AFM meronic state hosted by the free standing CrTe$_2$ layer. The spin-texture associated with the saddle point is illustrated in (b).} 
  \label{figure_3_r}
\end{figure}

Having established the existence of AFM meronic objects, we investigate their stability against thermal excitations utilizing a series of geodesic nudged elastic band (GNEB) simulations. 
Fig.~\ref{figure_3_r}(a) displays the minimum energy path for the collapse of the topological AFM state, which hosts an energy barrier of \SI{7.9}{\milli\electronvolt} (more details on the saddle point are shown in Fig.~\ref{figure_3_r}(b)).
This shows that these magnetic objects are metastable and should exist over a broad range of temperatures.

\subsection{Topological magnetic states in CrTe$_{2}$/XTe$_{2}$ heterobilayers}

Motivated by the intriguing magnetic behavior of the single CrTe$_{2}$ layer, we explore proximity-induced magnetic phases upon interfacing with various XTe$_{2}$ monolayers, X being a transition metal atom. Our systematic structural investigation of the different junctions enabled us to categorize the into three groups, as illustrated in Fig.~\ref{figure_1}(a):  
The first group hosts non-magnetic XTe$_2$ layers with a small lattice mismatch (less than 3\%) with CrTe$_{2}$ such as (Zr, Nb, Rh, Ni, Ti)Te$_{2}$, and this is the group that we focus on our work.
The second and third groups were disregarded since they have either a large lattice mismatch (more than 4\%) or are magnetic, which would lead to more complex proximity-induced effects to be explored in future studies.

\begin{figure}[htb]
\includegraphics[width=\columnwidth]{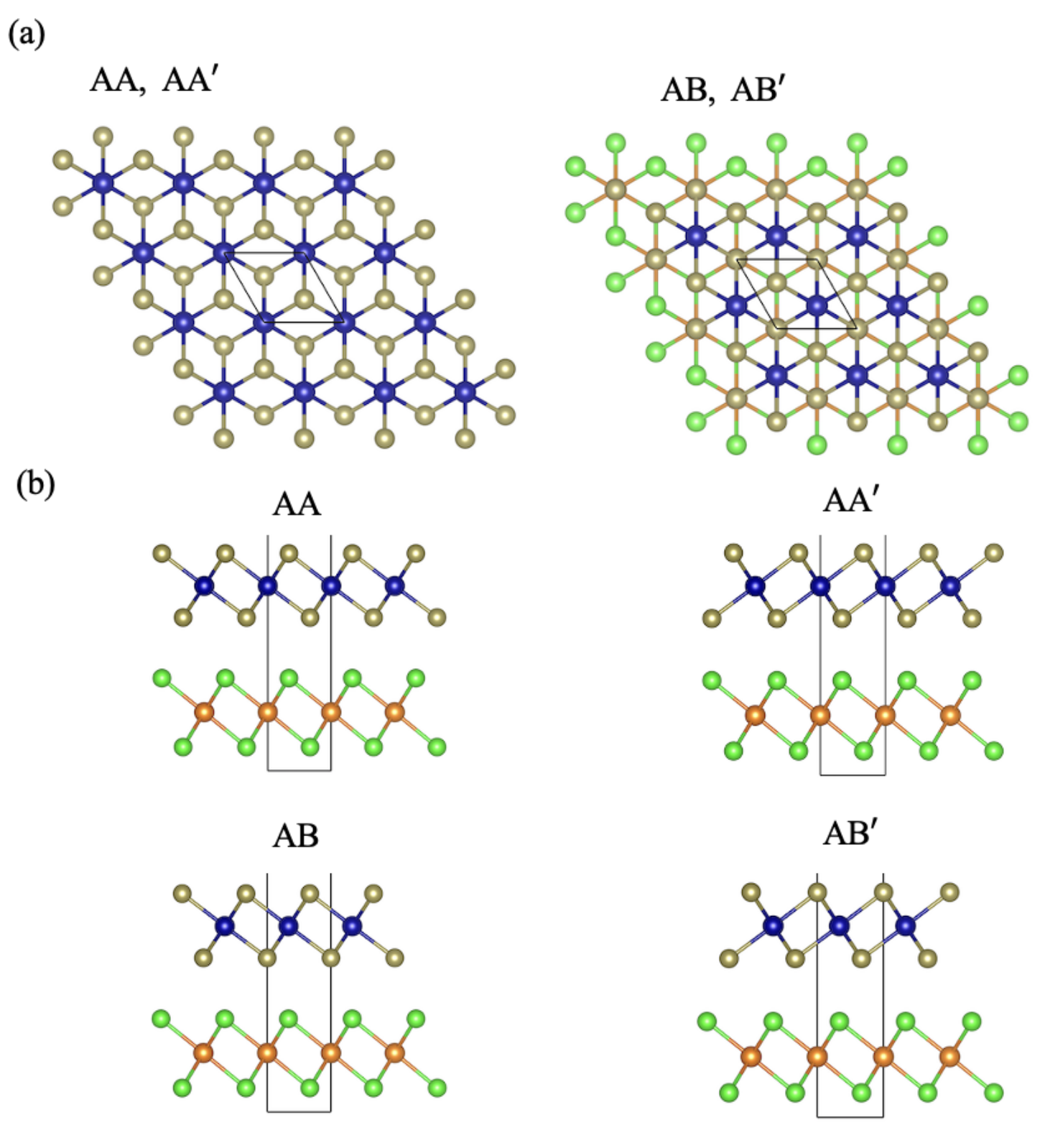}
   \caption{The different AA, AA$^\prime$, AB, and AB$^\prime$ stacking arrangements for CrTe$_{2}$/XTe$_{2}$ heterobilayers. (a) Top and (b) side views.} 
       \label{figure_3}
\end{figure}

The heterobilayers made of CrTe$_{2}$ and nonmagnetic XTe$_{2}$  were built assuming four different stacking (AA, AA$^\prime$, AB, and AB$^\prime$) as illustrated in Fig.~\ref{figure_3}.
In the AA stacking, which is the ground state (see Table~\ref{table:energy}), Cr is vertically aligned with the transition metal X, and the Te layers forming the interface are shifted with respect to each other, while in the AA$^\prime$ arrangement they are stacked on top of each other.
In the AB and AB$^\prime$ structures Cr and X are no longer vertically aligned, and the two structures are distinguished by the stacking arrangement of the Te layers at the interface.
In the following we focus our investigation on the AA stacking.

\begin{table*}[tb]
\centering
\caption{\label{table:energy} Total energy difference between various stacking orders and the ground state AA structure
for all CrTe$_{2}$/XTe$_{2}$ heterobilayers. Energies in meV.}
\begin{ruledtabular}
\bgroup
\def\arraystretch{1.5}
\begin{tabular*}{\textwidth}{l @{\extracolsep{\fill}} rrrrrr}
\multicolumn{7}{c}{CrTe$_{2}$ on top of} \\ \cline{2-7} 
Stacking & TiTe$_{2}$& NbTe$_{2}$& TaTe$_{2}$ & NiTe$_{2}$ &ZrTe$_{2}$& RhTe$_{2}$ \\
\hline
AA$^\prime$ & 105.3 & 126.9 & 125.9 & 149.1 & 130.1 &185.5 \\
AB & 30.3 & 27.3 & 26.4 &  103.6 & 60.5 & 120.8\\
AB$^\prime$ &  9.9 & 4.6 & 8.2 &  4.8 & 23.7 & 61.6\\
\end{tabular*}
\egroup
\end{ruledtabular}
\end{table*}

Table~\ref{table:para} shows the lattice parameters, including lattice constant and interlayer distance, for the AA stacking order.
It is clear that  these parameters vary significantly depending on the transition metal in the Te-based layers.
The lattice constants can be grouped around two values: $\sim \SI{3.7}{\angstrom}$ for CrTe$_{2}$/(Ti, Nb, Ta)Te$_{2}$ heterobilayer, which is close to the lattice constant of the free standing CrTe2 monolayer, and $\sim \SI{3.8}{\angstrom}$ in CrTe$_{2}$/(Zr, Ni, Rh)Te$_{2}$ heterobilayer, which is similar to the value of the bulk CrTe$_{2}$ lattice constant and results from the strain created at the interface.

\begin{table*}[tb]
\centering
\caption{\label{table:para} Lattice constant $a$, interlayer distance $h$, magnetic anisotropy energies ($K$) and Dzyaloshinskii-Moriya interaction ($D$) of the  CrTe$_{2}$/XTe$_{2}$ heterobilayers in the AA stacking. As a reference the free standing CrTe$_2$ has a lattice constant of \SI{3.71}{\angstrom} with a magnetic anisotropy of \SI{1.40}{\milli\electronvolt} and zero DMI.}
\begin{ruledtabular}
\bgroup
\def\arraystretch{1.5}
\begin{tabular*}{\textwidth}{c @{\extracolsep{\fill}} rrrrrr}
\multicolumn{7}{c}{CrTe$_{2}$ on top of} \\ \cline{2-7} 
 & TiTe$_{2}$& NbTe$_{2}$& TaTe$_{2}$ & NiTe$_{2}$ &ZrTe$_{2}$& RhTe$_{2}$ \\
\hline
$a$ (\AA) &3.73 & 3.70 & 3.70 & 3.81 & 3.82 &3.79 \\ 
$h$ (\AA) &3.76 & 3.65 & 3.74 &  3.46 & 3.75 & 3.50 \\
$K$ (meV) & 0.95& 0.90 & 0.94 & 0.90 & 0.61 & 1.70 \\
$|\mathbf{D}_1|$ (meV) &0.21& 0.23 & 0.27 & 0.30 & 0.30 & 0.45 \\
 $D_1^z$ (meV) & 0.10 &  0.15 & 0.22 & 0.05 & 0.06 & 0.06  \\
$|\mathbf{D}_2|$ (meV) & 0.48 & 0.37 & 0.48 & 0.60& 0.61 & 0.71\\
 $D_2^z$ (meV) &  0.15 &  0.18 &  0.20 &  0.01 &  0.01 &  0.12 \\
$|\mathbf{D}_3|$ (meV) & 0.38& 0.27 & 0.30 & 0.31 & 0.32 & 0.35 \\
 $D_3^z$ (meV) & 0.05 &  0.05 & 0.05 &  0.02 &  0.02 &  0.80 \\
\end{tabular*}
\egroup
\end{ruledtabular}
\end{table*}

\begin{figure}[htb]
\includegraphics[width=\columnwidth]{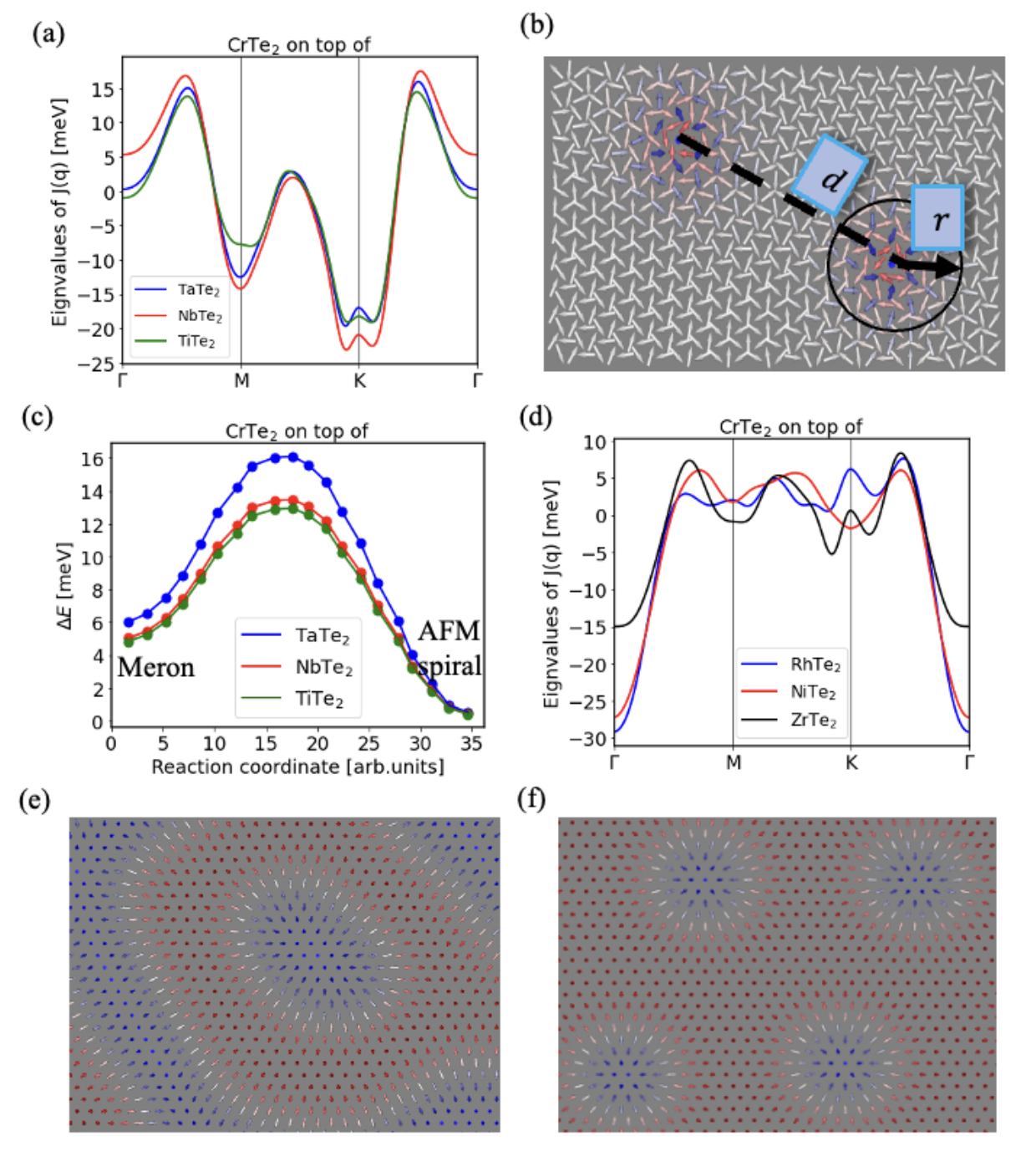}
   \caption{ Topological magnetic textures in heterobilayers. In (a,b,c) and (d,e,f) results related respectively to CrTe$_{2}$/(Ti,Nb,Ta)Te$_{2}$ and CrTe$_{2}$/(Zr, Ni, Rh )Te$_{2}$ heterobilayers. (a,d) Eigenvalues of the Fourier-transformed exchange interactions as a function of q.
   (b) An example of an AFM meronic texture indicating the radius $r$ of each meron and the distance $d$ between the meronic partners.
   (c) Energy path for the collapse of a single pair of AFM merons in the heterobilayers. 
   (e) Zero-field skyrmionic-like magnetic state in the heterobilayers. 
   (f) Skyrmionic lattice formed upon application of a magnetic field of \SI{6}{\tesla}. } 
       \label{figure_4}
\end{figure}

Now we turn to the analysis of the magnetic properties of all heterobilayers in the AA stacking.
In the first scenario, where the interfacing Te-based layer contains  Ta, Nb, or Ti, the Heisenberg exchange interactions induce a frustrated spin spiraling state with an energy minimum close to the N\'eel state (Fig.~\ref{figure_4}(a)). Once interactions induced by the spin-orbit interaction included, AFM multimeronic spin-textures emerge similar to the free-standing case (Fig.~\ref{figure_4}(b)).  Adding the substrate layer breaks the inversion symmetry of the CrTe$_{2}$ monolayer and introduces the DMI, where the $z$-component of the of the DM-vector for the first and second nearest neighbors favors in-plane rotations of the magnetic moments and so is compatible with the underlying in-plane magnetic anisotropy (Table~\ref{table:para}).
This enhances the stability of the AFM topological objects, as can be identified by the increased energy barrier illustrated in Fig.~\ref{figure_4}(c). 
Therefore, the heterobilayer with the largest $z$-component of the DMI has the highest energy barrier.
The energy barrier for the bilayers with TiTe$_{2}$, NbTe$_2$ and TaTe$_2$ is \SI{0.2}{\milli\electronvolt}, \SI{0.4}{\milli\electronvolt} and \SI{2.2}{\milli\electronvolt} higher than the one for free-standing CrTe$_2$, respectively, and the increased stability correlates with the increase in the magnitude of the DM interaction going from Ti to Nb to Ta, as listed in Table~\ref{table:para}.
We note that the MAE is roughly constant for all investigated interfaces ($\sim\SI{0.9}{\milli\electronvolt}$).
Interestingly, the radius $r$ of each of the merons and the distance $d$ between the them (illustrated in Fig.~\ref{figure_4}(b)) show opposite trends.
The radius increases for the heterobilayers, with values of \SI{2.6}{\nano\meter} (TiTe$_2$), \SI{2.7}{\nano\meter} (NbTe$_2$) and \SI{2.8}{\nano\meter} (TaTe$_2$) larger than the one for the free-standing CrTe$_2$ (\SI{2.4}{\nano\meter}), and follows the increase of the out-of-plane DMI.
Conversely, the distance between the two merons is progressively reduced: $d = \SI{34.8}{\nano\meter}$ for free-standing CrTe$_2$ and 34.4, 34.3 and \SI{34.0}{\nano\meter} once it is interfaced with TiTe$_2$, NbTe$_2$ and TaTe$_2$, respectively.

\begin{figure}[htb]
\includegraphics[width=\columnwidth]{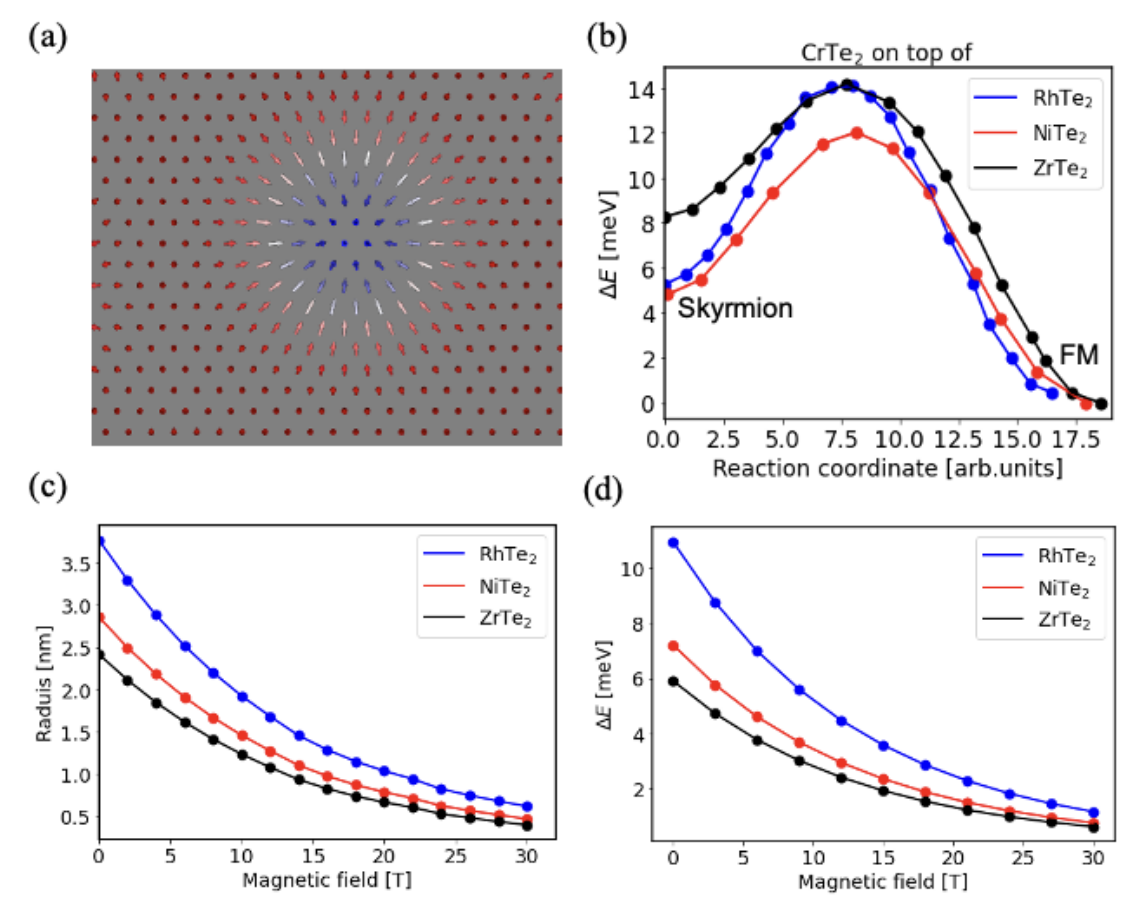}
   \caption{Single skyrmion in CrTe$_{2}$/(Zr, Ni, Rh )Te$_{2}$ heterobilayers. (a) Single skyrmion selected to calculate energy barriers and explore its stability.
   (b) Energy path for the collapse of the single skyrmion.
   (c) The radius of a single skyrmion and energy barriers, obtained with GNEB simulations, as a function of magnetic field.}
       \label{figure_5_r}
\end{figure}

In the second scenario, CrTe$_2$ is interfaced with Te-based layers hosting either Zr, Ni or Rh.
As mentioned before, these layers impose a lattice strain on CrTe$_2$ that switches the magnetic ground state from AFM to FM based on the Heisenberg exchange interactions (see energy minimum in Fig.~\ref{figure_4}(d)).
Once DMI and MAE are taken into account (see values in Table~\ref{table:para}), N\'eel-type skyrmionic domains form in zero magnetic field as shown in Fig.~\ref{figure_4}(e). 
For the calculation of energy barriers we select isolated skyrmions, resulting in the spin texture shown in Fig.~\ref{figure_5_r}(d), revealing that they are metastable.
A magnetic field larger than \SI{6}{\tesla} transforms the skyrmionic domain state to a triangular lattice of skyrmions, Fig.~\ref{figure_4}(f), which is more stable than the FM state by 3.1, 2.8 and \SI{2.2}{\milli\electronvolt} for the Rh, Ni and Zr-based heterobilayers, respectively.
Next we explore the thermal stability of an isolated skyrmion in a FM background, for which GNEB simulations  led to the energy barriers plotted in Fig.~\ref{figure_5_r}(b).
The barrier that has to be overcome to allow the skyrmion to relax to the FM state increases going from Ni (\SI{6}{\milli\electronvolt}) to Zr (\SI{7}{\milli\electronvolt}) to the Rh-based bilayer (\SI{8}{\milli\electronvolt}), which decreases with increasing the magnetic field as shown in Fig~\ref{figure_5_r}(d).
In this case the MAE depends more strongly on the nature of the heterobilayer (e.g., it is almost twice as large for Rh than for Ni), and hence has a stronger influence on the energy barrier and also on the skyrmion radius.
A larger energy difference between the skyrmion and the FM state corresponds to a larger skyrmion radius, and as expected they all shrink in size when the magnitude of the applied magnetic field is increased (Fig.~\ref{figure_5_r}(c)), disappearing above about \SI{30}{\tesla}.
In contrast, applying an external magnetic field to the AFM merons found for the Ti, Nb and Ta-based heterobilayers has no noticeable effect on their radius, but the underlying total topological charge changes at approximately \SI{12}{\tesla} from $\pm$1 to 0 (see Table~\ref{table:mag} for more details).

\begin{table}[htb]
\caption{\label{table:mag} Effect of the magnetic field on the topological charge of the AFM multi-meronic spin-texture.
Shown are the total topological charge $T$ and how it arises from the contributions of the individual sublattices $t_i$.}
\begin{ruledtabular}
\bgroup
\def\arraystretch{1.5}
\begin{tabular*}{\columnwidth}{l @{\extracolsep{\fill}} rr}
\multicolumn{2}{c}{Topological charges $T$ $(t_\alpha, t_\beta, t_\gamma)$} \\ \cline{1-2}
$B=\SI{0}{\tesla}$ & $B=\SI{12}{\tesla}$  \\
\hline
0 (0, 0, 0), (0, -1, 1) &0  (0, 0, 0), (0, -1, 1)  \\ 
1 (1, 0, 0), (1, -1, 1) &0 (0, 0, 0), (0, -1, 1)  \\
-1 (-1, 0, 0), (-1, -1, 1) &0 (0, 0, 0), (0, -1, 1) \\
\end{tabular*}
\egroup
\end{ruledtabular}
\end{table}

\section{Conclusions}
In summary, we demonstrated that the monolayer of CrTe$_{2}$ in the 1T phase can host various topological magnetic states once interfaced with non-magnetic Te-based layers hosting transition metals by using a combination of density functional theory calculations and atomistic spin dynamics.
The scrutinized quantum materials were pre-selected to be within an acceptable range of lattice mismatching, which makes the considered atomic structures realistic and hence make our predictions of the magnetic properties more reliable.

Our main finding is the emergence of a new type of antiferromagnetic topological state consisting of hexamer-meronic spin-texture in a magnetically frustrated environment characterizing the free-standing CrTe$_{2}$ as well as the CrTe$_{2}$/(Ta, Nb, Ti)Te$_{2}$ heterobilayers.
This magnetic state forms in a rich set of pair combinations of merons and antimerons, with each pair living in one of the three antiferromagnetic sublattices.
By constructing the vdW bilayers, inversion symmetry is broken, which gives rise to a $z$-component of the Dzyaloshinskii-Moriya interaction that enhances the stability of these novel meronic textures.

Intriguingly, when CrTe$_2$ is instead proximitized with (Zr, Rh, Ni)Te$_{2}$ layers it displays ferromagnetic behavior, which is imposed by the interface-induced strain, hosting spin-spirals as well as ferromagnetic skyrmions which are both enabled by the DMI.
These results provide a potential explanation for the anomalous Hall effect identified in the CrTe$_{2}$/ZrTe$_{2}$ heterostructures\cite{zrte2}.
We note that a recent work predicts the formation of skyrmions in CrTe$_2$/WTe$_2$ bilayer~\cite{crte2-wte2}, which is an interface that we disregarded because of the large lattice mismatch.

\begin{figure}[htb]
\centering
    \includegraphics[width=\columnwidth]{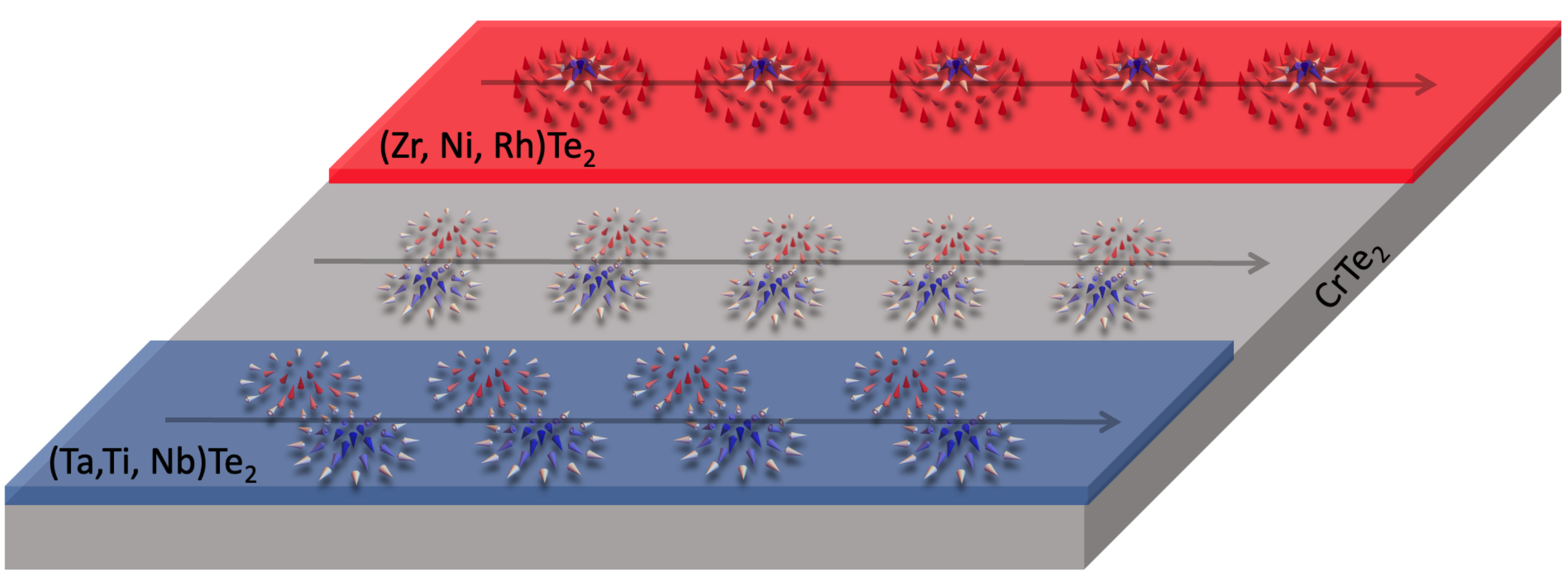}
   \caption{Potential technological device concept combining 2D CrTe$_2$ with other 2D layers, such as (Ta, Nb, Ti)Te$_2$ which promote antiferromagnetic merons as topological magnetic defects, and (Zr, Rh, Ni)Te$_2$ layers that promote ferromagnetic skyrmions.
   The various types of topological spin textures can then be injected from one device region to another, and driven using applied spin currents or thermal gradients, for instance.} 
       \label{figure_7}
\end{figure}

Overall, our work highlights CrTe$_2$ as a promising 2D layer for further exploration of proximity-induced topological magnetism enabled by its strong magneto-elastic coupling~\cite{Abuawwad}.
Our findings suggest the possibility of engineering the size and stability of the underlying topological spin-textures by modifying the nature of the interfacing 2D material.
More importantly, we anticipate that besides the fundamental importance of identifying the frustrated antiferromagnetic multi-meronic textures, patching the same 2D material such as CrTe$_2$ with distinct 2D layers such as those unveiled in this work, favoring either ferromagnetic skyrmions of antiferromagnetic merons, can be useful constituents of information technology devices.
We envisage, for instance, their potential application for the ultimate control and transport of dissimilar topological objects to carry information in well designed regions of multiple 2D vdW heterojunctions, as schematically depicted in Fig.~\ref{figure_7}.

\begin{acknowledgments}
We acknowledge fruitful discussions with Amal Aldarawsheh, Jose Martinez-Castro and Markus Ternes. This work was supported by the Federal Ministry of Education and Research of Germany in the framework
of the Palestinian-German Science Bridge (BMBF grant number 01DH16027).  We acknowledge funding provided by the Priority Programmes SPP 2244 "2D Materials Physics of van der Waals heterobilayer"  (project LO 1659/7-1) and SPP 2137 “Skyrmionics” (Projects LO 1659/8-1) of the Deutsche Forschungsgemeinschaft (DFG).
We acknowledge the computing time granted by the JARA-HPC Vergabegremium and VSR commission on the supercomputer JURECA at Forschungszentrum Jülich~\cite{jureca}. 
The work of MdSD made use of computational support by CoSeC, the Computational Science Centre for Research Communities, through CCP9.
\end{acknowledgments}

\bibliography{references}

\end{document}